\documentclass[fleqn,usenatbib]{mnras}

\usepackage{newtxtext,newtxmath}
\usepackage[T1]{fontenc}

\DeclareRobustCommand{\VAN}[3]{#2}
\let\VANthebibliography\thebibliography
\def\thebibliography{\DeclareRobustCommand{\VAN}[3]{##3}\VANthebibliography}

\usepackage{graphicx}	% Including figure files
\usepackage{amsmath}	% Advanced maths commands
\def \bexrb {BeXRB}
\def \bexrbs {BeXRBs}
\def \scubed {S-CUBED}
\def \swift {Swift}

\def \smcxtwo {SMC~X-2}
\title[SMC X-2 2022]{The 2022 super-Eddington outburst of the source SMC X-2}

\author[M. J. Coe et al.]
{M. ~J. Coe,$^{1}$\thanks{E-mail: mjcoe@soton.ac.uk}
J.~A. Kennea$^{2}$,
I.~M. Monageng$^{3,4}$,
L.J. Townsend$^{3,5}$,
D.A.H. Buckley$^{3,4}$,  
  \newauthor
M. Williams$^{2}$,
A. Udalski$^{6}$
and
P.~A. Evans$^{7}$\\
$^{1}$Physics \& Astronomy, The University of Southampton, Southampton, SO17 1BJ, UK\\
$^{2}$Department of Astronomy and Astrophysics, The Pennsylvania State University, 525 Davey Lab, University Park, PA 16802, USA\\
$^{3}$South African Astronomical Observatory, P.O Box 9, Observatory, 7935, Cape Town, South Africa\\
$^{4}$Department of Astronomy, University of Cape Town, Private Bag X3, Rondebosch 7701, South Africa\\
$^{5}$Southern African Large Telescope, Cape Town, South Africa\\
$^{6}$Astronomical Observatory, University of Warsaw, Al. Ujazdowskie 4, 00-478 Warszawa, Poland\\
$^{7}$University of Leicester, X-ray and Observational Astronomy Research Group, School of Physics \& Astronomy, University Road, Leicester LE1 7RH, UK.\\
}

\date{Accepted XXX. Received YYY; in original form ZZZ}

\pubyear{2022}

\begin{document}
\label{firstpage}
\pagerange{\pageref{firstpage}--\pageref{lastpage}}
\maketitle

\begin{abstract}

SMC X-2 exhibits X-ray outburst behaviour that makes it one of the most luminous X-ray sources in the Small Magellanic Cloud. In the last decade it has undergone two such massive outbursts - in 2015 and 2022. The first outburst is well reported in the literature, but the 2022 event has yet to be fully described and discussed. That is the goal of this paper. In particular, the \textcolor{black}{post-peak} characteristics of the two events are compared. This reveals clear similarities in decay profiles, believed to be related to different accretion mechanisms occurring at different times as the outbursts evolve. The H$\alpha$ emission line indicates that the Be disc undergoes complex structural variability, with evidence of warping as a result of its interaction with the neutron star. The detailed observations reported here will be important for modelling such interactions in this kind of binary systems. 

\end{abstract}


%%%%%%%%%%%%%%%%%%%%%%%%%%%%%%%%%%%%%%%%%%%%%%%%%%

%%%%%%%%%%%%%%%%% BODY OF PAPER %%%%%%%%%%%%%%%%%%

\section{Introduction}

SMC X-2 is a high-mass X-ray binary system located in the Small Magellanic Cloud (SMC), a satellite galaxy of the Milky Way. The system consists of a neutron star and a massive companion star, which are orbiting each other in a close binary system.

The neutron star in SMC X-2 is a highly compact object with a mass approximately 1.4 times that of the sun, but a radius of only a few kilometers. The companion star is a massive OB-type star, which is a type of hot, luminous star that emits strong Balmer lines in its spectrum due to the presence of a circumstellar disk of gas.

The binary system is slightly eccentric (e=0.07), with an orbital period of 18.4 days \citep{townsend2011}. As the two stars orbit each other, material from the companion star's circumstellar disc is pulled towards the neutron star due to its strong gravitational field. This material may form a transient accretion disk around the neutron star, which emits X-rays as it is heated up and undergoes gravitational interactions with the neutron star.

SMC X-2 is an important object of study for astrophysicists because it is a high-mass X-ray binary system (HMXB), which is a class of objects that are thought to be provide significant sources of X-rays in the universe. The system is also notable for exhibiting a range of complex X-ray variability. These include regular 2.37s pulsations from the neutron star, as well as more erratic variability that is likely caused by changes in the circumstellar disc size and hence accretion rate onto the neutron star.

The first reported super-Eddington outburst for \smcxtwo\ was in 2000 \citep{2001corbet}. It then underwent an extended period of quiescence until 15 years later when the system exhibited another major X-ray outburst reaching X-ray luminosities well in excess of $10^{38}$ erg/s. Observational reports and follow-up discussions of this particular outburst may be found in the following papers: \cite{2017lutovinov}, \cite{2016li}, \cite{2015foto}, \cite{2016palombara}, \cite{2015kennea}, \textcolor{black}{\cite{Roy2022} and \cite{Jaisawal2023}}. 7 years later the source underwent the next major X-ray outburst which is reported in this paper. Though this 2022 outburst didn't quite reach the same X-ray peak luminosity as the 2015 event it nonetheless again exceeded $10^{38}$ erg/s and hence represents a substantial occurrence. In this work we report a multiwaveband study of this event and make important comparisons with the earlier 2015 X-ray outburst.

\section{Observations}

\subsection{Swift X-ray Observations}

The \scubed{} project \citep{kennea2018} started regular observations using the Swift X-ray Telescope (XRT; \citealt{burrows05}) of the Small Magellanic Cloud in July 2016 in order to watch for X-ray outbursts from \bexrb\ ~systems. As part of this project \smcxtwo\ has been observed approximately weekly in the 0.3-10~keV band, although the majority of these observations are only 60s exposure.  We note that the previous outburst of \smcxtwo\ was in 2015, and was tracked extensively by \swift\ through Target of Opportunity (TOO) observations, however this outburst was over before the \scubed{} project began. 

On June 14th, 2022 data from  \scubed{} showed a point source at the location of \smcxtwo{} detected in 38s of Photon Counting (PC) mode XRT data, significantly above all previous upper limits found in \scubed{} monitoring of this source \cite{2022kennea}. This indicated that \smcxtwo was likely entering a new outburst phase. 

Due to the detection of this new outburst observations were requested through the Swift Target of Opportunity (TOO) program, consisting of initially daily, followed by every 2 day observations with a requested exposure time of 1ks per exposure. Observations were taken in a combination of PC and Windowed Timing (WT) modes for XRT, based on predicted X-ray brightness. These TOO observations continued until \smcxtwo{} was no longer detectable by \swift{}, with the final observation taken on September 19th, 2022. Alongside these TOO observations, \scubed{} regular monitoring of the source continued.

\begin{figure}
	\includegraphics[width=8cm,angle=-0]{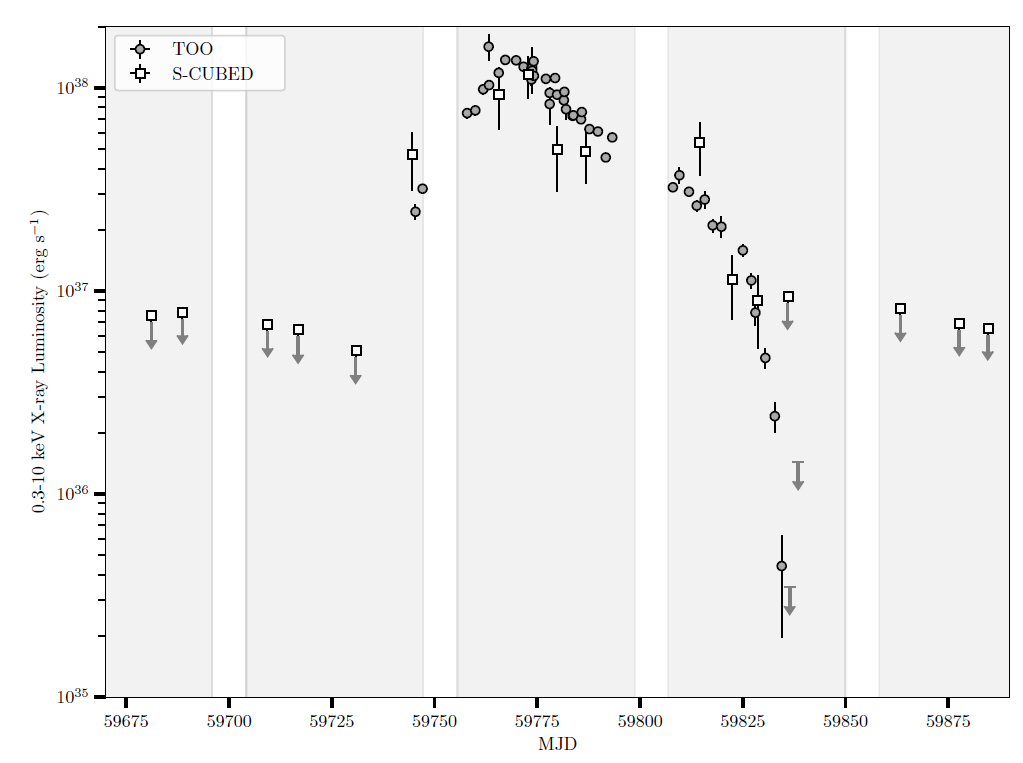}
    \caption{Combined \scubed{} and TOO X-ray light-curve of SMC X-2, showing the \textcolor{black}{2022} outburst period. Grey background regions show periods of time when the source was  observable by \swift{}, white gaps are observing gaps due to spacecraft constraints. \textcolor{black}{The similarity of the 2022 outburst profile can be compared to the 2015 outburst profile (see Fig.~1 from \citet{2017lutovinov}).}}
    \label{fig:xlc}
\end{figure}

A table of TOO observations is given in Table~\ref{tab:obs} and the resulting measurements displayed in Fig.~\ref{fig:xlc}.

\begin{figure}
	\includegraphics[width=8cm,angle=-0]{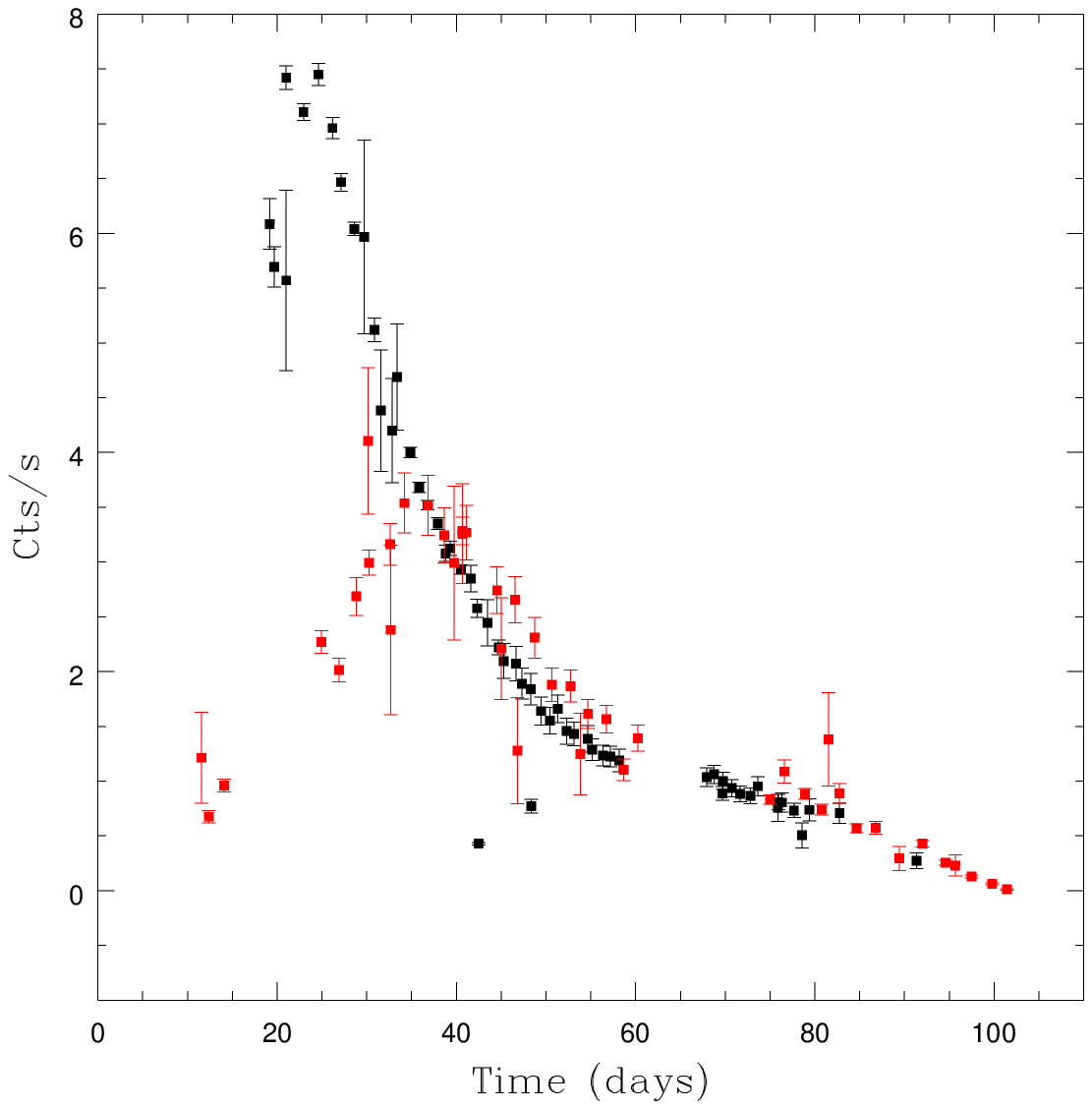}
    \caption{The X-ray profiles (XRT counts/s in 0.3-10~keV range) of the 2015 outburst (black) overlaid on top of the 2022 outburst discussed in this paper (red). \textcolor{black}{The 2022 data is the same as that presented in Fig.~\ref{fig:xlc}, but on a linear scale and in flux units.}}
    \label{fig:outbursts}
\end{figure}

Fig.~\ref{fig:outbursts} shows the X-ray decay profiles of the 2022 outburst overlaid on the profile for the 2015 one. To produce this merged plot the 2015 data  timings have been arbitrarily adjusted by -57280d and the 2022 data by -59740d. \textcolor{black}{There has been no attempt to precisely match the data sets, but rather a small offset has been chosen so that both data sets can be clearly seen to follow similar decay patterns}. Though the peak fluxes differ by a factor of two, the decay profiles are extremely similar. In both cases there is a distinct change in the decay slope around the 60d mark in the figure. These decay profiles are discussed further below.

\subsection{UV and optical photometry}

\begin{figure*}
	\includegraphics[width=18cm,angle=-0]{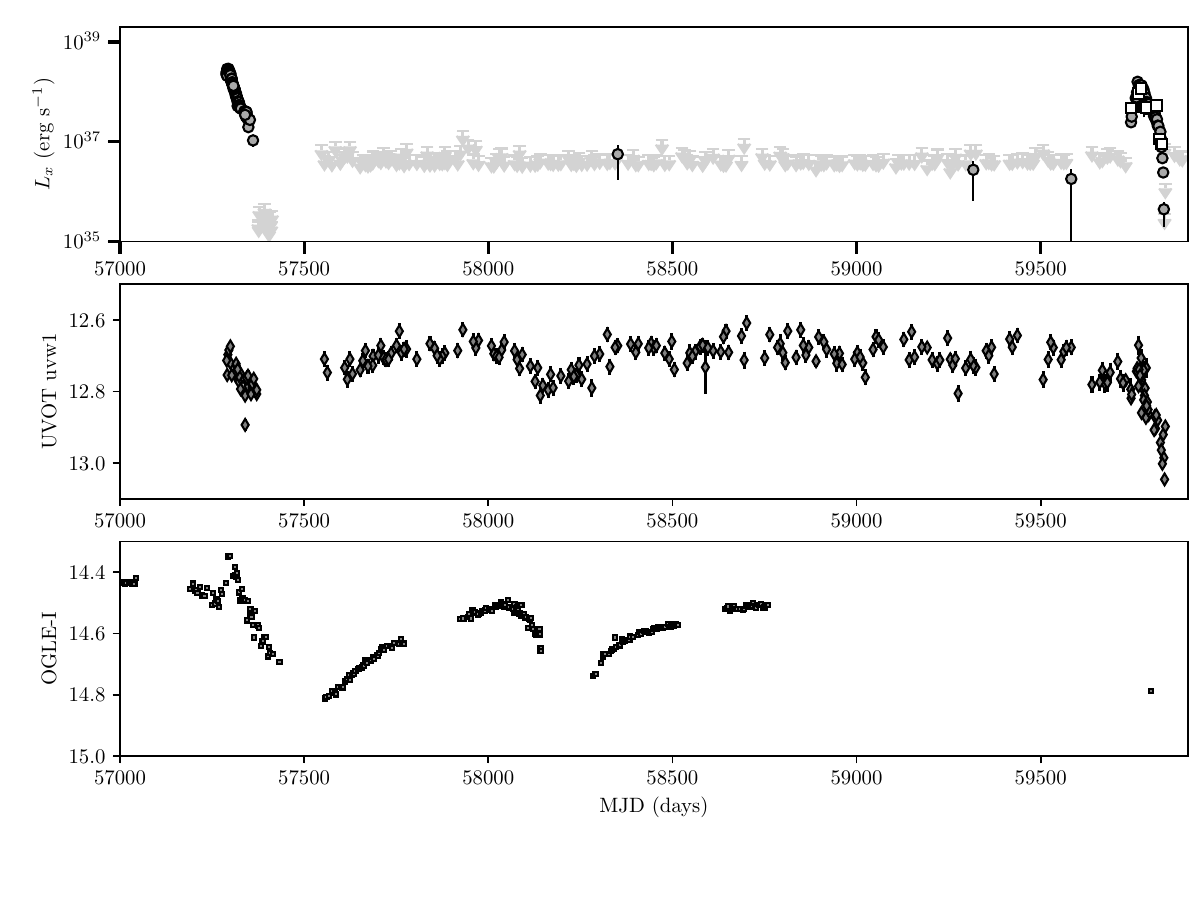}
    \caption{8 year time history of Swift XRT 0.3-10~keV (top panel), UVOT uvw1 filter (middle panel), with OGLE I-band lightcurve plotted for comparison (bottom panel). \textcolor{black}{The 2015 outburst covers the period MJD 57300 - 57450, and the 2022 outburst MJD 59750 - 58400.}}
    \label{fig:threeband}
\end{figure*}

The UltraViolet and Optical Telescope (UVOT) uvw1 filter (2600\AA) data were analyzed using the Swift ftools available in HEASOFT version 6.30. Images with multiple snapshots were summed with uvotimsum. Photometry was then performed using uvotsource with a source region of radius 5'' and a larger source-free background region. Images that were not aligned with the World Coordinate System were first aspect corrected using uvotunicorr before being analyzed as previously described. The resulting lightcurve is shown in Fig.~\ref{fig:threeband}

The OGLE project \citep{Udalski2015} provides long term I-band photometry with a cadence of 1-3 days. I band data were collected on the optical counterpart to \smcxtwo\ as part of the OGLE IV project from JD 2455346 till JD 2458759. The OGLE identification for this object is SMC720.17.50. These results are also shown in Fig.~\ref{fig:threeband}.

\subsection{Optical spectroscopy}

\begin{figure}
	\includegraphics[width=8cm,angle=-0]{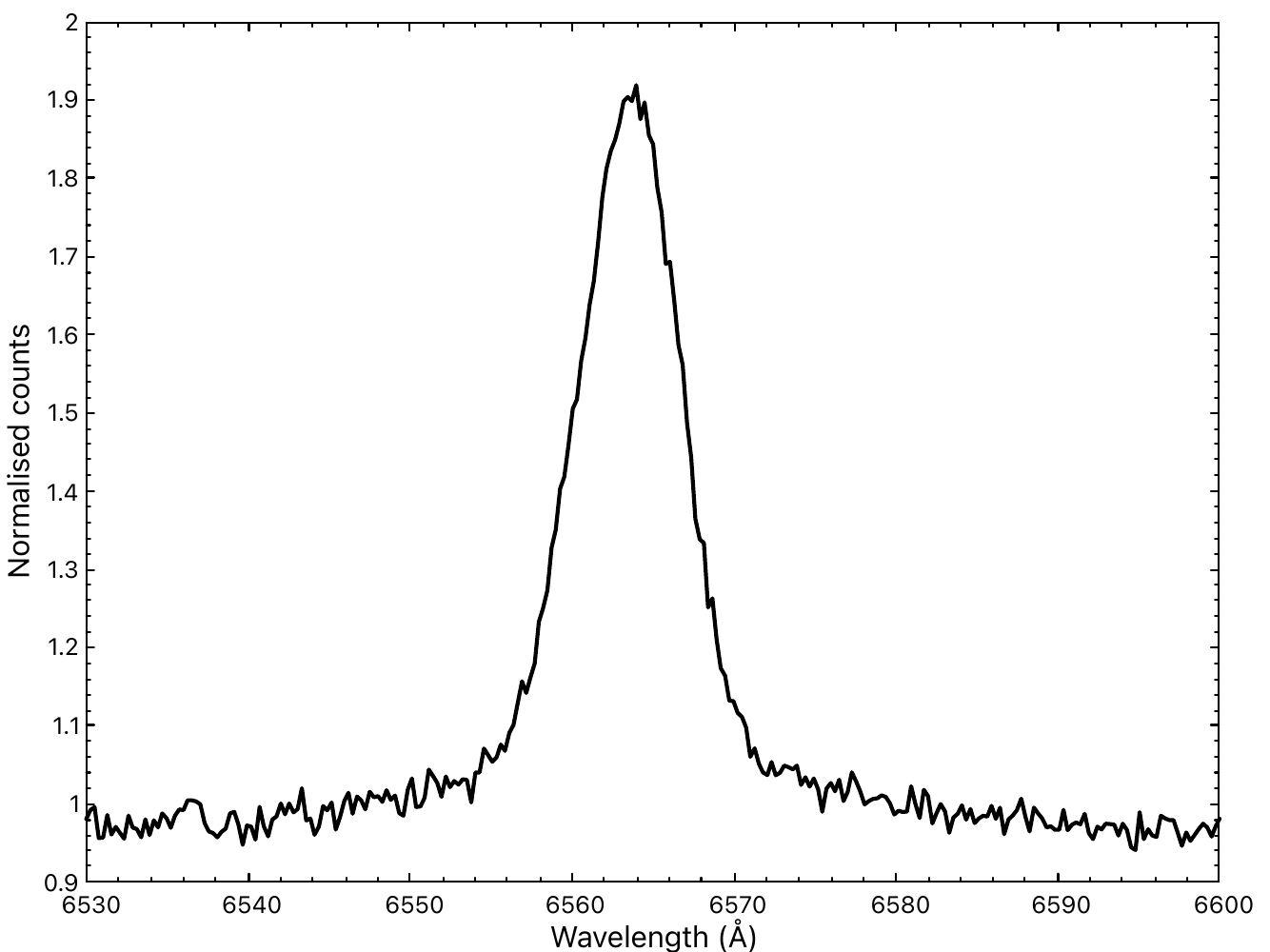}
    \caption{An example of the H$\alpha$ emission line in \smcxtwo\ from an RSS observation obtained on 07-08-2022 (MJD59798.07).}
    \label{fig:H_a_eg}
\end{figure}

The optical counterpart to \smcxtwo\ was observed with the Southern African Large Telescope (SALT; \citealt{2006SPIE.6267E..0ZB}) using the Robert Stobie Spectrograph (RSS; \citealt{2003SPIE.4841.1463B,2003SPIE.4841.1634K}) and the High-Resolution Spectrograph (HRS; \citealt{2010SPIE.7735E..4FB,Bramall2012,2014SPIE.9147E..6TC}). The RSS observations were performed using various grating settings: PG0900, PG1800 and PG2300. A summary of the settings is provided in Table~\ref{tab:gratings}. The primary reductions (which comprise overscan correction, bias subtraction, gain correction and amplifier cross-talk corrections) were performed using the SALT pipeline \citep{2012ascl.soft07010C}. The remaining data reduction steps (arc line identification, background subtraction and 1D spectrum extraction) were executed using \textsc{iraf}\footnote{Image Reduction and Analysis Facility: iraf.noao.edu}.\\
The HRS observations were carried out in low-resolution mode ($R \sim 14 000$), where exposure times of 1200~s were used. A wavelength coverage of 3750--8790~\AA~was obtained. The SALT pipeline was used for the primary reductions, while the remaining reduction steps ((background subtraction, identification of arc lines, blaze function removal, and merging of the orders) were done using the \textsc{midas feros} \citep{1999ASPC..188..331S} and \textsc{echelle} \citep{1992ESOC...41..177B} packages. The full details of the data reduction process are described in \cite{2016MNRAS.459.3068K}.\\
A log of the observations and the inferred H$\alpha$ equivalent width (EW) measurements are shown in Table~\ref{tab:SALT_log}. An example of the H$\alpha$ emission line profile is shown in Fig.~\ref{fig:H_a_eg}. 

\begin{table}
	\centering
	\caption{A summary of the settings for the SALT RSS observations}
	%\resizebox{\columnwidth}{!}
	\label{tab:gratings}
    \setlength\tabcolsep{0.8pt}
	\begin{tabular}{cccc} % four columns, left alignment for each
		\hline\hline
		 & Grating  & Wavelength & Exposure     \\ [-3pt]
		Grating &  angle &  range (\AA) & time (s)   \\
		\hline
		PG0900 & 12.5/14.75 & $4070-7000$  & 120/600 \\
		PG1800 & 36.5 & $5900-7200$  & 240  \\
		PG2300 & 48.875 & $6100-6900$  & 1800  \\
		\hline
	\end{tabular}
\end{table}

However, obtaining a reliable assessment of the true strength of the H$\alpha$ EW for \smcxtwo\ is difficult. The optical counterpart has been shown to have a very close, similar star on the sky, separated by just $\sim$1 arcsecond \citep{1979Murdin}. Those authors report that the nearby star is an O7 type object, which means it probably has an H$\alpha$ EW of +(3.0--3.4)\AA  ~\citep{1974conti}. The vast majority of the SALT spectra were obtained whilst the seeing conditions were 1-2 arcsec, therefore it is probable that there is some contamination of the determined H$\alpha$ EW values for \smcxtwo\   by up to this amount. If the O7 star does not have a circumstellar disc, and hence does not exhibit variable H$\alpha$ EW, then the estimates presented here of the H$\alpha$ EW for \smcxtwo\ could be misrepresented in magnitude by up to +3\AA. So it cannot be assumed that any apparent changes in this parameter from \smcxtwo\ are real if the changes are less than $\sim$3\AA.\\

\subsection{MeerKAT}
We obtained three observations of \smcxtwo\ with MeerKAT during its outburst. The observations were performed at a central frequency of 1.28~GHz, bandwidth of 856~MHz and an integration time of 8~s. Each observation consisted of a total of 60-minute scans on target. J0408-6545 was used as a primary calibrator, where it was observed at the start and end of each scan. The secondary calibrator, J0252-7104, was observed in 2-minute scans for every 20-minute on the target. The data were processed using the OxKAT reduction routines \citep{2020ascl.soft09003H}\footnote{https://github.com/IanHeywood/oxkat}, which performs averaging, application of standard bandpass and gain corrections, and flagging. The imaging was executed using \textsc{wsclean}, after which a self-calibration cycle was performed. All three observations resulted in non-detections at the optical position of \smcxtwo\ . A log of the 3$\sigma$ upper limits is given in Table~\ref{tab:MK}. The radio emission during peak outburst is believed to be due to the launch of an accretion-powered jet in these systems (e.g. \citealt{vdEinjden2021,vdEinjden2022}). However, the current sensitivity of instruments is inadequate to detect jet emission at extragalactic distances. The increased sensitivity from future radio telescopes in the Southern hemisphere is expected to probe the jet-launching mechanism.

\begin{table}
\begin{tabular}{llc}
\hline
       Date & MJD         &  Radio flux density ($\mu$Jy)        \\
\hline
2022-06-19	&	59749.5 & < 38.4  \\
2022-07-10	&	59770.5 & < 38.6  \\
2022-07-18	&	59778.5 & < 34.9  \\

\hline
\end{tabular}
\caption{MeerKAT (1.28~GHz) 3$\sigma$ upper limits.}
 \label{tab:MK}
\end{table}

\section{Discussion}
\subsection{X-ray decay profiles}

Fig.~\ref{fig:outbursts} shows the combined X-ray decay profiles of the 2015 and 2022 outbursts. Simple linear best fits have been applied to the two phases of the decay with a break at 40d (TJD 57320 for the 2015 outburst and TJD 52513 for the 2022 event). This break corresponds to a XRT count rate of $\sim$1.5 cts/s, or a luminosity in the SMC of $\sim5 \times 10^{37}$ erg/s. At this point the the decay profile slope completely changes and flattens by a factor of $\sim$6.5. \textcolor{black}{Other simple models were explored to fit the decay section of the S-CUBED data such as an exponential fit. Such a fit works well for the first section of the data, but after the 40d point the data indicate a flatter decay profile. Clearly more sophisticated modelling beyond the scope of this paper are required.}

There is also a suggestion of an even steeper decay rate during the first 10-15d of the 2015 outburst, that switch occurring at a luminosity of $\sim1.5 \times 10^{38}$ erg/s. \textcolor{black}{Unfortunately the 2022 outburst did not reach such high luminosities so we cannot comment on such a steep profile component.} 

These changes in decay profile rates almost certainly indicate significant changes in the mode of accretion on to the neutron star. \cite{2017lutovinov} studied the changing decay profile of the 2015 outburst and concluded that there was a change of accretion mode to the propeller mode \citep{1975is} at a luminosity of $4 \times 10^{36}$ erg/s which corresponds to an XRT count rate of $\sim$0.1 cts/s on Fig.~\ref{fig:outbursts}. This is at the very end of the outburst, at which point they \textcolor{black}{show that the X-ray flux dropped significantly. There is no doubt that this is the end of the outburst but it is at the limit of the sensitivity of the Swift XRT telescope. Important to their model are the upper limits set immediately after the final positive detection since they help define the shape of the final decay phase. They set the upper limit for any detection as low as $2 \times 10^{34}$ erg/s. Unfortunately the data presented here are less sensitive and are only able to set a limit an order of magnitude higher - see Fig.~\ref{fig:xlc}.}

\cite{2016Tsygankov} investigated the switching on/off of the propeller effect in two other \bexrbs - 4U 0115+63 and V 0332+53. In both cases they interpreted the tail end of the outburst decays to indicate an accretion transition around $(1-2) \times 10^{36}$ erg/s. It is interesting to see the visual difference that presenting the data on a logarithmic scale (e.g. Fig.~\ref{fig:xlc}) as compared to a linear scale (e.g. Fig.~\ref{fig:outbursts}). Plus the difficulties of fitting models that rely on the limiting sensitivity of the telescope - see, for example, the results presented by \cite{2016Tsygankov} on the source V 0332+53 (their Figure 2). The manner of the data presented here focuses strongly on the times of very significant source detection, and the resulting change in decay rates seen at $\sim5 \times 10^{37}$ erg/s. This is unlikely to be a switch to the propeller regime since the flux does not drop dramatically, but it certainly indicates that something significant has happened that deserves further modelling investigations beyond the scope of this paper. Finally, it is worth noting that the binary period for this system is 18.4d with a slight eccentricity of 0.07 \citep{townsend2011}, so throughout the observational period being discussed here ($\sim$70d) the distance between the neutron star and the fuel source, the circumstellar disc, is constantly varying. So, inevitably, the accretion environment being experienced by the neutron star is also constantly changing. This factor also needs to be included in any models.

\subsection{Optical photometric variability}
\subsubsection{Long-term UVOT uvw1 and I band lightcurves}
The evolution of the UVOT uvw1 and OGLE I-band brightness taken over a period of $\sim$8 years is shown in the bottom 2 panels of Fig.~\ref{fig:threeband}. The lightcurves reveal long-term variability that is an indicator of structural changes happening in the inner and outer regions of the disc.\\
Both the uvw1 and I-band magnitudes generally show correlated changes with respect to the X-ray lightcurve. \textcolor{black}{This is particularly evident during the decline of both the 2015 and 2022 X-ray outbursts. The I-band lightcurve shows a sharp spike at the onset of the 2015 X-ray outburst, which suggests that the neutron star distorted the outer parts of the disc which resulted in the matter being accreted, leading to enhanced X-ray activity. This behaviour was discussed by \cite{Roy2022}. Similarly, an enhanced spike in the uvw1 lightcuve is observed at the start of the 2022 X-ray outburst followed by a drop in flux as the X-ray outburst decays}. During the decay of the 2015 X-ray outburst, the I-band brightness is seen to decrease at a faster rate than the uvw1. This is unsurprising since the I-band magnitude is a tracer of the more extended, outer regions of the disc compared to the uvw1 magnitude which measures the changes in the inner disc regions. \textcolor{black}{It is often noted that the longer wavelength filters (i.e. the I-band) undergo larger amplitude changes compared to those at the shorter wavelengths in BeXBs (e.g. \citealt{Reig2015,Coe2022})}. This behaviour indicates global disruption of the disc from the outer to inner parts as a result of the interaction of the disc with the neutron star. \textcolor{black}{A noticeable feature of Fig.~\ref{fig:threeband} during the simultaneous uvw1 and I-band observations is the time lag in reaching peak brightness in the two bands between MJD57900 and MJD58100. The innermost region of the disc reaches peak brightness $\sim$150 days before the outermost parts of the disc, indicative of the time taken for the effects of a mass ejection to reach the outer limits of the disc.}

\subsubsection{(uvw1-I) colour variability}
\begin{figure}
	\includegraphics[width=8cm,angle=-0]{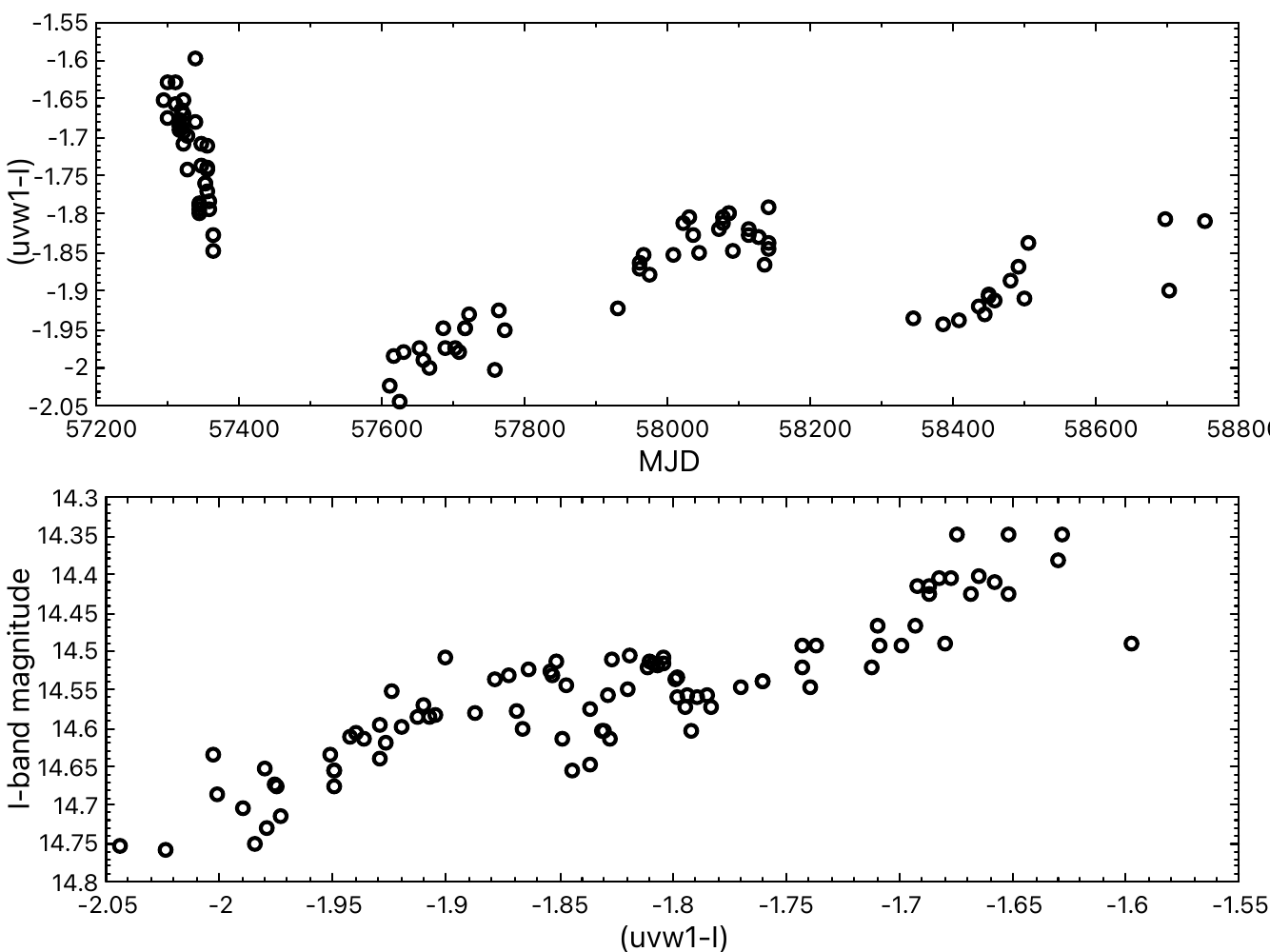}
    \caption{\textcolor{black}{Top panel: The evolution of the $(uvw1-I) - I$ colour. Bottom panel: $(uvw1-I) - I$ colour–magnitude diagram.}}
    \label{fig:simultaneous}
\end{figure}

\textcolor{black}{The top panel of Fig.~\ref{fig:simultaneous} shows the evolution of the (uvw1-I) colour for the period where there are simultaneous uvw1- and I-band observations. The colour is a proxy for the disc temperature, with the blue and red colours indicating hotter and cooler temperatures, respectively. During the period between MJD57300 and MJD57400 the overall (uvw1-I) colour evolves to lower (bluer) values, indicating an overall temperature increase. This is the period when the overall brightness was undergoing a decline (Fig.~\ref{fig:threeband}) during the decay of the 2015 X-ray outburst. This suggests that the outer, cooler parts of the disc were lost due to the neutron star interaction. The overall trend of the (uvw1-I) colour between MJD57600 and MJD58800 is increasing (the colour is reddening) since this is the period when the disc is increasing in size and there is therefore a larger contribution of cooler temperatures as the outer regions extend outwards.}\\

\textcolor{black}{As circumstellar discs in BeXBs grow in size they generally show an excess flux and their red continuum increases. The reddening of the system with disc growth is attributed to the outer parts of the disc being cooler than the inner parts \citep{Harmanec1983}. When viewed at low/intermediate angles, a correlation between the brightness and colour (i.e. temperature) would be observed. Discs that are viewed at an edge-on orientation (high inclination angles) would exhibit an anti-correlation between the brightness and colour. In this geometry, as the disc grows in size the overall brightness of the Be disc/star system decreases since the flaring disc blocks out some of the light from the Be star \citep{Harmanec1983,Rajoelimanana2011,Reig2015}.} Fig.~\ref{fig:simultaneous} (bottom panel) shows the $(uvw1-I) - I$ colour-magnitude plot for the optical counterpart of SMC X-2. A strong correlation is observed between the $(uvw1-I)$ colour and the brightness, which indicates an intermediate/low inclination angle of the disc. This is corroborated by the morphology of the H$\alpha$ emission line in Fig.~\ref{fig:H_a_eg}, which does not show a shell profile with a central depression that goes below the continuum, as is conventionally seen in highly-inclined disc orientations \citep{Reig2015}.

\subsection{Optical spectroscopic variability}
\label{sec:residual}

Using Kepler's laws we estimate the semi-major axis and periastron passage of the neutron star, assuming the canonical neutron star mass ($M_X \sim 1.4$~M$_\odot$) and a Be star mass of $M_\star \sim 24$~M$_\odot$ based on its spectral type \citep{Straizys1981}. The orbital parameters ($P = 18.38$~d and $ecc = 0.07$) derived by \citet{townsend2011} were used for the calculations. Using these parameters we find values of $86.20\pm0.10$~R$_\odot$ and \textcolor{black}{$80\pm2$~R$_\odot$} for the semi-major axis and periastron passage, respectively.

Using the method of \cite{1989Ap&SS.161...61H}, we estimate the size of the disc from the largest measured equivalent width (13.15~\AA), which hopefully has the least contamination from the nearby O7 star. It gives a maximum radius that is larger than the periastron distance of the neutron star orbit ($111\pm7$~R$_\odot$). \\

\begin{figure*}
	\includegraphics[width=18cm,angle=-0]{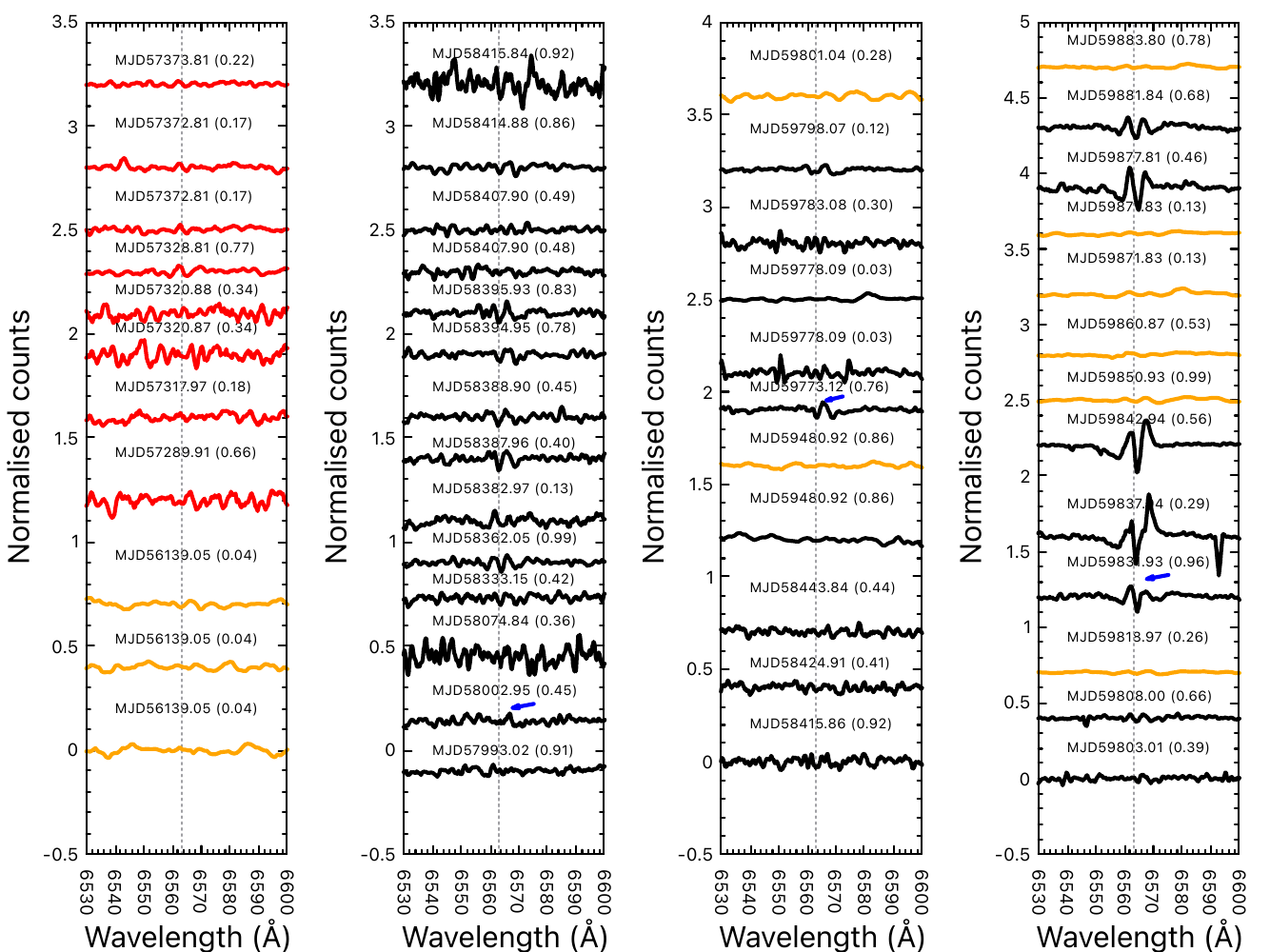}
    \caption{Voigt-subtracted H$\alpha$ residual spectra. The dashed vertical line in each panel represents the rest lab wavelength of the H$\alpha$ line. The modified Julian dates and orbital phases were calculated using the ephemeris from \citet{townsend2011}. \textcolor{black}{The blue arrows indicate the epochs of the onset of the redshifted peaks as discussed in Section~\ref{sec:residual}}. The different colours indicate the different gratings used for the observations: PG0900 (orange), PG1800 (red), and PG2300 (black).}
    \label{fig:Voigt_subtract}
\end{figure*}

In Fig.~\ref{fig:Voigt_subtract} we show the evolution of residual H$\alpha$ emission line spectra. To obtain these, we first created a model emission line by fitting a Voigt function to each spectrum and then subtracted the observed spectra from the model. Since the Voigt model approximates emission from an axisymmetric circumstellar disc, the resultant residual is a measure of deviation from axisymmetry. For the analysis we only consider observations taken with the PG2300 grating, as the resolution from the PG0900 and PG1800 is insufficient to study the morphological behaviour of the disc. \textcolor{black}{Evidence of deviation from symmetry is seen before the 2022 X-ray outburst, at MJD58002.95, as there is a residual structure that shows up as a peak on the redshifted side of the line. This residual profile is seen to develop into a double-peak profile in the subsequent observations until MJD58395.93. The redshifted peak is seen to develop again during the peak of the 2022 X-ray outburst (MJD59773.12).}
The subtle variability in the residual profile continued as the outburst declined until MJD59831.93 when a pronounced double-peaked structure is seen that persists until the end of the outburst. This is a possible indicator of a complex non-uniform structure at the outer parts of the disc. \\
The semi-major axis can be calculated using the orbital parameters above, which when combined with the projected semi-major axis, $a_X \sin i_{\textrm{orbit}} \sim 32$ R$_\odot$ \citep{townsend2011}, can be used to estimate the inclination angle ($i_{\textrm{orbit}} \sim 22^\circ$) of the plane of the neutron star orbit. \textcolor{black}{The disc radius can be estimated using the double-peaked H$\alpha$ emission from the relation \citep{Huang1972}}:
\begin{equation}
R_{CS} = \frac{GM_{\star} \sin^2i_{\textrm{disc}}} {(0.5 \Delta V)^2},
\label{eq:Huang}
\end{equation}
\textcolor{black}{where:
\begin{itemize}
    \item $G$ - gravitational constant
    \item $M_{\star}$ - mass of the Be star
    \item $i_{\textrm{disc}}$ - inclination angle of the disc
    \item $\Delta V$ - peak separation 
\end{itemize}
The peak separation, $\Delta V$, is measured by fitting a two-peak Voigt profile to the double-peaked emission lines and using the central wavelengths of the fitted peaks in the calculation. }Assuming that the residual component of the disc extends out to the outer edges, up to a radius of around $100$~R$_\odot$, using Equation~\ref{eq:Huang} and the spread of measured peak separations ($\sim 250 - 300$~km/s) gives a range of $35 - 45^\circ$ for the outer edge inclination angles. This misalignment between the plane of the orbit of the neutron star and the plane of the Be disc can result in the warping of the outer disc edges. This is especially true when the circumstellar disc has grown to a size approaching that of the neutron star orbit.

\section{Conclusions}
 \textcolor{black}{The source \smcxtwo\ has shown three major outbursts in the last two decades - in 2000, 2015 and 2022. The one reported here which occurred in 2022 though not quite as X-ray bright as the 2015 event,  is, however, well-documented at multiple wavelengths. At its peak it exceeded the traditional Eddington Limit with an X-ray luminosity of over $10^{38}$ erg/s, and triggered supporting optical \& UV observations, both spectroscopic and photometric. It is only by integrating all these different multi-wavelength data sets together that valuable insights are obtained into both the details of the accretion physics of neutron stars, as well as the dynamic behaviour pattern of the companion Be star that triggered the whole event. The evidence that the Be star triggered the 2022 X-ray outburst is the contemporaneous sharp rise in the UV flux reported here - indicative of a major mass ejection from the star. This is the fuel for accretion onto the neutron star and the resulting X-ray outburst. Such behaviour is well established in other BeXRB systems - see, for example, \cite{2020kennea} and \cite{2019monageng}.}

 \textcolor{black}{The characteristics of the 2015 and 2022 decay profiles are compared revealing clear similarities, including a significant change in the rate of decay believed to be related to differing accretion mechanisms occurring at different times. The H$\alpha$ emission line profiles indicates that the Be disc undergoes complex structural variability during the outburst, with evidence of warping as a result of its interaction with the neutron star. The detailed observations reported here will be important for more sophisticated modelling of such interactions in BeXRB systems, perhaps using Smooth Particle Hydrodynamics (see, for example, \cite{2019Brown}).}

 \textcolor{black}{}

\section*{Acknowledgements}

 JAK and MW acknowledge support from NASA Grant NAS5-00136. PAE acknowledges UKSA support. LJT is supported by the SALT Foundation and the South African NRF. DAHB is supported by the South African NRF. IMM is supported by the UCT VC 2030 Future Leaders Programme and the South African NRF.

%%%%%%%%%%%%%%%%%%%%%%%%%%%%%%%%%%%%%%%%%%%%%%%%%%
\section*{Data Availability}

All X-ray data are freely available from the NASA Swift archives. The OGLE optical data in this article will be shared on any reasonable request to Andrzej Udalski of the OGLE project.



\bibliographystyle{mnras}
\bibliography{references} % if your bibtex file is called example.bib

%\begin{thebibliography}{99}
%\end{thebibliography}

%%%%%%%%%%%%%%%%%%%%%%%%%%%%%%%%%%%%%%%%%%%%%%%%%%

%%%%%%%%%%%%%%%%% APPENDICES %%%%%%%%%%%%%%%%%%%%%

%\appendix


%%%%%%%%%%%%%%%%%%%%%%%%%%%%%%%%%%%%%%%%%%%%%%%%%%
\appendix

\section{Tables of X-ray and optical observations}

\begin{table}
\centering
    \begin{tabular}{ccc}
    \hline
    Obs Date & Flux & Flux \\
    MJD&cts/s&error\\
    \hline
59744.57	&	0.87	&	0.35	\\
59745.37	&	0.63	&	0.05	\\
59747.09	&	0.82	&	0.04	\\
59757.93	&	1.91	&	0.10	\\
59759.92	&	2.00	&	0.11	\\
59761.84	&	2.55	&	0.15	\\
59763.16	&	4.11	&	0.59	\\
59763.27	&	2.64	&	0.10	\\
59765.61	&	3.13	&	0.19	\\
59765.68	&	2.13	&	0.72	\\
59767.21	&	3.53	&	0.07	\\
59769.85	&	3.52	&	0.07	\\
59771.65	&	3.24	&	0.07	\\
59772.77	&	2.91	&	0.70	\\
59773.67	&	3.21	&	0.38	\\
59773.67	&	3.00	&	0.12	\\
59774.13	&	3.27	&	0.26	\\
59777.55	&	2.74	&	0.14	\\
59778.00	&	2.21	&	0.47	\\
59779.57	&	2.66	&	0.13	\\
59779.80	&	1.28	&	0.49	\\
59781.72	&	2.31	&	0.09	\\
59783.67	&	1.87	&	0.20	\\
59785.73	&	1.86	&	0.05	\\
59786.82	&	1.45	&	0.37	\\
59787.69	&	1.61	&	0.13	\\
59789.74	&	1.56	&	0.05	\\
59791.65	&	1.09	&	0.15	\\
59793.24	&	1.41	&	0.12	\\
59807.99	&	0.83	&	0.04	\\
59809.59	&	0.96	&	0.09	\\
59811.90	&	0.78	&	0.04	\\
59813.79	&	0.67	&	0.04	\\
59814.55	&	1.05	&	0.37	\\
59815.74	&	0.74	&	0.07	\\
59817.67	&	0.54	&	0.04	\\
59819.79	&	0.53	&	0.06	\\
59825.03	&	0.41	&	0.03	\\
59827.58	&	0.26	&	0.02	\\
59830.49	&	0.12	&	0.01	\\
59832.78	&	0.06	&	0.01	\\
\hline
\end{tabular}
\caption{\label{tab:obs} \textcolor{black}{Summary of all Swift XRT detections of SMC X-2 during the 2022 outburst}}
\end{table}

\begin{table}
\begin{tabular}{lllcl}
\hline
       Date & MJD         &  EW (\AA)           &   Binary phase & RSS grating\\
\hline

2012-07-31	&	56139.05	& -9.96	 $\pm$ 0.25	& 0.04	& PG0900	  \\ 
2012-07-31	&	56139.06	& -10.54	 $\pm$   1.0	& 0.04	& PG0900  \\ 
2012-07-31	&	56139.06	& -9.83	 $\pm$ 	 0.69	& 0.04	& PG0900  \\ 
2015-09-24	&	57289.91	& -10.40 $\pm$	 0.47	& 0.66	& PG1800  \\ 
2015-10-22	&	57317.97	& -10.33 $\pm$	 0.28	& 0.18	& PG1800  \\ 
2015-10-25	&	57320.87	& -12.51 $\pm$	 0.55	& 0.34	& PG1800  \\ 
2015-10-25	&	57320.88	& -10.97 $\pm$	 0.52	& 0.34	& PG1800  \\ 
2015-11-02	&	57328.81	& -9.57 $\pm$	 0.59	& 0.77	& PG1800  \\ 
2015-12-16	&	57372.81	& -10.67 $\pm$	 0.37	& 0.17	& PG1800  \\ 
2015-12-16	&	57372.81	& -10.58 $\pm$	 0.44	& 0.17	& PG1800  \\ 
2015-12-17	&	57373.81	& -10.88 $\pm$	 0.32	& 0.22	& PG1800  \\ 
2017-08-28	&	57993.02	& -10.29 $\pm$	 0.45	& 0.91	& PG2300  \\ 
2017-09-06	&	58002.95	& -10.05 $\pm$	 0.36	& 0.45	& PG2300  \\ 
2017-11-17	&	58074.84	& -10.12 $\pm$	 0.73	& 0.36	& PG2300  \\ 
2018-08-03	&	58333.15	& -9.74 $\pm$	 0.28	& 0.42	& PG2300  \\ 
2018-09-01	&	58362.05	& -10.19 $\pm$	 0.32	& 0.99	& PG2300  \\ 
2018-09-21	&	58382.97	& -9.89 $\pm$	 0.56	& 0.13	& PG2300  \\ 
2018-09-26	&	58387.96	& -9.65 $\pm$	 0.21	& 0.40	& PG2300  \\ 
2018-09-27	&	58388.90	& -9.07 $\pm$	 0.38	& 0.45	& PG2300  \\ 
2018-10-03	&	58394.95	& -9.55 $\pm$	 0.29	& 0.78	& PG2300  \\ 
2018-10-04	&	58395.93	& -9.11 $\pm$	 0.53	& 0.83	& PG2300  \\ 
2018-10-16	&	58407.90	& -8.81 $\pm$	 0.17	& 0.48	& PG2300  \\ 
2018-10-16	&	58407.95	& -8.77 $\pm$	 0.56	& 0.49	& PG2300  \\ 
2018-10-23	&	58414.88	& -9.79 $\pm$	 0.19	& 0.86	& PG2300  \\ 
2018-10-24	&	58415.84	& -6.88 $\pm$	 0.45	& 0.92	& PG2300  \\ 

2018-10-24	&	58415.86	& -8.88 $\pm$	 0.18	& 0.92	& PG2300 \\ 
2018-11-02	&	58424.91	& -8.29 $\pm$	 0.32	& 0.41	& PG2300 \\ 
2018-11-21	&	58443.84	& -9.44 $\pm$	 0.49	& 0.44	& PG2300 \\ 
2021-09-23	&	59480.92	& -13.15 $\pm$	 1.16	& 0.86	& PG0900 \\ 
2021-09-23	&	59480.92	& -10.95 $\pm$	 0.84	& 0.86	& PG0900\\ 
2022-07-13	&	59773.12	& -9.46 $\pm$	 0.66	& 0.76	& PG2300 \\ 
2022-07-18	&	59778.09	& -10.26 $\pm$	 0.45	& 0.03	& PG2300 \\ 
2022-07-18	&	59778.12	& -11.72 $\pm$	 0.20	& 0.03	& PG2300 \\ 
2022-07-22	&	59783.08	& -10.29 $\pm$	 0.45	& 0.30	& PG2300 \\ 
2022-08-07	&	59798.07	& -10.00 $\pm$	 0.36	& 0.12	& PG2300 \\ 
2022-08-10	&	59801.04	& -11.75 $\pm$	 0.43	& 0.28	& PG0900 \\ 

2022-08-12	&	59803.01	& -10.35 $\pm$	 0.59 & 0.39	& PG2300  \\ 
2022-08-17	&	59808.00	& -9.98 $\pm$	 0.42 & 0.66	& PG2300  \\ 
2022-08-27	&	59818.97	& -11.47 $\pm$	 0.32 & 0.26	& PG0900  \\ 
2022-09-09	&	59831.93	& -10.79 $\pm$	 0.43 & 0.96	& PG2300  \\ 
2022-09-15	&	59837.94	& -8.52 $\pm$	 0.38 & 0.29	& PG2300  \\ 
2022-09-20	&	59842.94	& -8.40 $\pm$	 0.37 & 0.56	& PG2300  \\ 
2022-09-28	&	59850.93	& -11.15 $\pm$	 0.19 & 0.99	& PG0900  \\ 
2022-10-08	&	59860.87	& -11.16 $\pm$	 0.41 & 0.53	& PG0900  \\ 
2022-10-19	&	59871.83	& -10.80 $\pm$	 0.45 & 0.13	& PG0900  \\ 
2022-10-19	&	59871.87	& -10.04 $\pm$	 0.33 & 0.13	& PG0900  \\ 
2022-10-25	&	59877.81	& -4.05 $\pm$	 0.02 & 0.46	& PG2300  \\ 
2022-10-29	&	59881.84	& -9.32 $\pm$	 0.32 & 0.68	& PG2300  \\ 
2022-10-31	&	59883.80	& -11.14 $\pm$	 0.30 & 0.78	& PG0900  \\ 
2022-11-17	&	59900.87 & -7.09 $\pm$	 0.38	&  0.71	& HRS \\

\hline
\end{tabular}
\caption{A log of the SALT observations. The H$\alpha$ equivalent width measurements are recorded and the binary phase determined from the ephemeris in \citet{townsend2011}.}. 
 \label{tab:SALT_log}
\end{table}

% Don't change these lines
\bsp	% typesetting comment
\label{lastpage}
\end{document}